# Bending sound around sharp corners without using topological edge states

Liting Wu[1], Mourad Oudich[2,3], Wenkang Cao[1], Haolin Jiang[1], Cheng Zhang[1], Junchen Ke[1], Jin Yang[1], Yuanchen Deng[2], Qiang Cheng[1,*], Tiejun Cui[1,*] and Yun Jing[2,*]

[1]*State Key Laboratory of Millimeter Waves, Department of Radio Engineering, Southeast University, Nanjing 210096, People's Republic of China*

[2]*Department of Mechanical and Aerospace Engineering, North Carolina State University, Raleigh, North Carolina 27695, USA*

[3]*Universite de Lorraine, CNRS, Institut Jean Lamour, F-54000 Nancy, France*

Email: qiangcheng@seu.edu.cn, tjcui@seu.edu.cn, yjing2@ncsu.edu

**Routing and guiding acoustic waves around sharp corners without backscattering losses is of great interest in the acoustics community. Sonic crystals have been primarily utilized to design backscattering-immune waveguides. While conventional approaches use defects to guide waves, a considerably more sophisticated and robust approach was recently developed based on topological edge states. In this paper, we propose a radically different theoretical framework based on extremely anisotropic metamaterials for engineering backscattering-immune waveguides. We theoretically derived the exact condition for one-way wave propagation in zigzag paths, and identified a number of key advantages of the current design over topologically protected waveguides. While the theoretical underpinning is universal and is applicable to acoustic and electromagnetic waves, the experimental validation was conducted using spoof surface acoustic waves. The proposed metamaterial could open up new possibilities for wave manipulation**

**and lead to applications in on-chip devices and noise control.**

Routing and guiding acoustic waves without energy loss is a central topic in both fundamental and applied research of wave manipulation[1-9]. To achieve highly efficient routing of acoustic waves around sharp corners, the issue of acoustic energy loss due to intrinsic backscattering at the interface must be addressed. Utilizing linear defects in sonic crystals (SCs) to guide acoustic waves around sharp corners was first proposed by M. Torres *et. al* [10]. This design was later optimized by A. Khelif *et. al* [1] to obtain full-transmission guided waves by harnessing line defects. Alternatively, acoustic metamaterials (AMMs) can give rise to robust wave guiding by leveraging transformation acoustics or near-zero-density[8-9]. However, these methods are generally limited to a very narrow operating bandwidth or certain bending angles (usually 90° for defect-based methods). Recently, the field of robust acoustic wave guiding has enjoyed a strong revival, propelled mainly by the advent of nontrivial topological edge states discovered in SCs. A flurry of activity has been devoted to engineering topologically protected waveguides for backscattering-immune one-way acoustic propagation[11-20]. However, these waveguides are poorly coupled to the background media and SCs are wavelength-scaled structures that can become extremely bulky at low frequencies. The foregoing limitations underscore the importance of identifying alternative strategies for robust wave guiding. Additionally, while a majority of previous studies have focused on bulk acoustic waves, the present study addresses another important form of acoustic waves, *i.e.,* spoof surface acoustic waves (SSAWs), though the theory outlined in this paper is generic to acoustic and electromagnetic waves. To the best of our knowledge, backscattering-immune one-way propagation of SSAWs has not been experimentally observed before. This paper provides a rigorous analysis and demonstration using extremely anisotropic (EA-) AMMs for backscattering-immune one-way propagation of SSAWs.

AMMs are rationally designed materials composed of periodic subwavelength unit cells that can be used to manipulate sound waves in unprecedented ways [21-34]. Many promising applications, such as subwavelength imaging [29-30] and acoustic cloaking [31] have been demonstrated. EA-AMMs with large values of positive mass densities have been studied in recent years [32-34], and they are best known for their flat equal-frequency contours (EFCs). On the other hand, SSAWs, hosted by various artificial structures, such as one-dimensional (1D) corrugated rigid interfaces [35-39] and 2D corrugated surfaces/SCs [40-45], have gained attention in the acoustics community. Compared with conventional surface acoustic waves, which stem from the coupling between longitudinal and transverse waves in solids, SSAWs are guided along the interface between the background fluid medium (e.g., air) and a corrugated perfect rigid body (PRB), and that contributes to their unique applications in lab-on-a-chip devices, energy trapping, and sound focusing. An additional benefit of SSAWs is that they provide a fertile ground for investigating a myriad of wave phenomena since they hold a crucial advantage over bulk acoustic waves; the entire SSAW wave field can be "non-invasively" measured at a plane slightly above the PRB surface, providing a full metamaterial-acoustic interaction map.

In this study, we first examined a unique class of 2D EA-AMMs for guiding bulk acoustic waves. The homogenization theory was employed to analyze the acoustic characteristics derived from the strong density anisotropy. Several interesting features of these EA-AMMs were discovered through theoretical and numerical investigations. For instance, we discovered that the propagation route in this type of EA-AMM is independent of the angle of incidence, and unlike acoustic topological insulators or SCs hosting valley hall states, these materials can be efficiently coupled with the background media under a wide range of incident angles, beam widths, ad source locations. More importantly, the exact condition for backscattering-immune propagation in a zigzag path with sharp corners was rigorously established. The generality of the theory further enables us to implement the concept on a small size EA-AMM for SSAWs, where the entire wave field of interest is experimentally "mapped" and compared with the theory and simulation. All results unambiguously demonstrate robust wave guiding around sharp corners, without relying on the recently emerging paradigm of topological edge state.

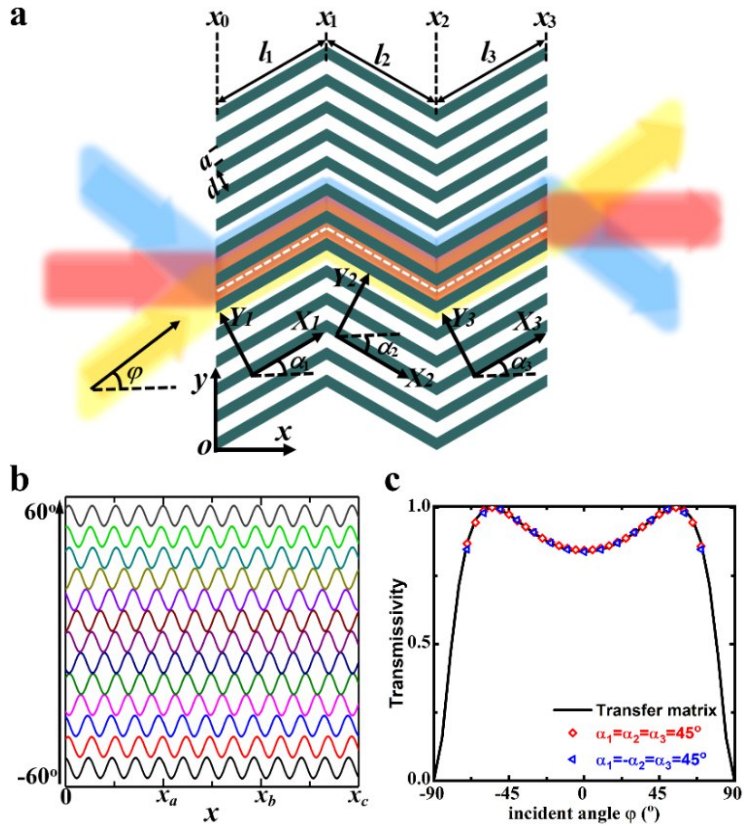

**Fig. 1 | Schematic illustration of an EA-AMM with a zigzag path hosting backscattering-immune sound propagation. a** Two-dimensional configuration of the EA-AMM with a zigzag path. **b** Simulated normalized acoustic pressure distributions at 5000 Hz, along the white dashed lines marked in **a**. The angle of incidence ranges from -60° to 60°, with 10° increments. **c** Simulated transmissivity $T$ for a straight waveguide with $α_1 = α_2 = α_3 = 45°$ (marked by red diamonds) and a zigzag waveguide with $α_1 = -α_2 = α_3 = 45°$ (marked by blue triangles), and the theoretical results calculated by the transfer matrix method. The simulation results presented here are real-structure simulation results.

Figure 1a illustrates a 2D EA-AMM composed of a periodic array of stitching parallelogram PRBs separated by air gaps. For better illustration, we define $(X_i, Y_i)$ ($i$=1, 2, 3) and $(x, y)$ as the local and global coordinate systems, respectively. $d$ and $a$ are the periodicity of the array and the width of the waveguide (air gap), respectively, and they are both in the subwavelength range, allowing only the fundamental acoustic mode in the waveguide. Based on the effective medium theory, the subwavelength array structure shown in Fig. 1a can be homogenized as a 2D continuous acoustic medium with effective anisotropic mass densities ($\rho_X$ and $\rho_Y$) and effective bulk modulus ($B$) in the local coordinate system [37]:

$$\rho_X = (d/a)\rho_0, \quad \rho_Y = \infty, \quad B = (d/a)B_0, \tag{1}$$

with $\rho_0$ and $B_0$ being the mass density and bulk modulus of air, respectively. To further characterize the metamaterial, the effective parameters are rewritten in the global coordinate system. The corresponding inverse mass density tensors $1/\rho_{xx}^i$, $1/\rho_{yy}^i$, $1/\rho_{xy}^i$ and $1/\rho_{yx}^i$, and the bulk modulus $B_i$ can be written as [46]

$$\frac{1}{\rho_{xx}^i} = \frac{\cos^2\alpha_i}{\rho_X^i} + \frac{\sin^2\alpha_i}{\rho_Y^i}, \tag{2}$$

$$\frac{1}{\rho_{yy}^i} = \frac{\cos^2\alpha_i}{\rho_Y^i} + \frac{\sin^2\alpha_i}{\rho_X^i}, \tag{3}$$

$$\frac{1}{\rho_{xy}^i} = \frac{1}{\rho_{yx}^i} = \sin\alpha_i \cos\alpha_i \left(\frac{1}{\rho_X^i} - \frac{1}{\rho_Y^i}\right), \tag{4}$$

$$B_i = B, \tag{5}$$

where $\alpha_i$ ($i$ = 1, 2, 3) are the angles of rotation of the local coordinate system relative to the global coordinate system (marked in Fig. 1a). It will be later shown that these three angles must be identical in magnitude to ensure backscattering-immune sound propagation. Here, we define that, when the coordinate system ($X_i$, $Y_i$) rotates anticlockwise, the angle of rotation $\alpha_i$ is positive; otherwise it is negative. With this model, the dispersion relation can be obtained and further utilized to derive the group velocity of the metamaterial in different regions (Supplementary Section 1), that is,

$$\vec{v}_g = \sqrt{\frac{B_0}{\rho_0}} \begin{pmatrix} \cos\alpha_i \\ \sin\alpha_i \end{pmatrix}. \tag{6}$$

It is evident that in such an AMM with $\rho_Y \to \infty$, the group velocity depends only on the angles of rotation $α_i$, and is independent of the angle of incidence $φ$. It should be noted that the group velocity $\vec{v}_g \equiv \nabla_{\vec{k}} ω(\vec{k})$ specifies the direction of energy flow for the wave, and is not necessarily parallel to the wave vector $\vec{k}$ (or the phase velocity) of the metamaterial [47]. Therefore, the direction of the energy flow can be predetermined by choosing the desired $α_i$.

Furthermore, the reflection coefficient $r_{ii'}$ and transmission coefficient $t_{ii'}$ from region $i$ to $i'$ ($i = i' \pm 1$) can be shown to satisfy (Supplementary Section 2)

$$|r_{ii'}| = \left| \frac{n_i/ρ_{xx}^i - n_{i'}/ρ_{xx}^{i'}}{n_i/ρ_{xx}^i + n_{i'}/ρ_{xx}^{i'}} \right|, \quad |t_{ii'}| = \left| \frac{2 n_{i'}/ρ_{xx}^{i'}}{n_i/ρ_{xx}^i + n_{i'}/ρ_{xx}^{i'}} \right|, \tag{7}$$

where $n_i \equiv \sqrt{ρ_{xx}^i c_0^2/B - (ρ_{xx}^i)^2 \sin^2 φ/ρ_X ρ_Y}$. This equation is instrumental as it describes the transmission efficiency around sharp corners for the proposed EA-AMM, which is governed by the angle of rotation $α_i$ ($ρ_{xx}^i$ contains $α_i$). On the basis of this equation, total transmission can be obtained by letting $|α_i| = |α_{i'}|$, giving rise to zero reflection. That is to say, for a zigzag path with $α_i = -α_{i'}$, backscattering-immune transmission in this EA-AMM can be attained. This is not a severe limitation though, since in theory any bending angle can be divided into two identical angles (See Supplementary Figure 20). Figure 1b illustrates the simulated normalized pressure field along the white dashed lines marked in Fig. 1a with different angles of incidence at 5000 Hz. Here we set $α_1 = -α_2 = α_3 = 45°$, $a = 3.2$ mm, and $d = 4$ mm, which are far smaller than the wavelength (wavelength is $λ = 58.6$ mm). The simulations are carried out by COMSOL Multiphysics. The magnitudes of the acoustic pressure in the three regions are virtually identical, demonstrating backscattering-immune sound propagation in the EA-AMM under a wide range of incident angles.

Next, we investigated the coupling between the background medium and the EA-AMM shown in Fig. 1a. The pressure reflectivity $R$ and transmissivity $T$ at the interface between air and the EA-AMM can be calculated by a transfer matrix method as follows (see Supplementary Section 3 for details)

$$R = \frac{(\xi - \frac{1}{\xi})^2 \sin^2(n_1 k \Delta x)}{(\xi + \frac{1}{\xi})^2 \sin^2(n_1 k \Delta x) + 4\cos^2(n_1 k \Delta x)},$$

$$T = \frac{4}{(\xi + \frac{1}{\xi})^2 \sin^2(n_1 k \Delta x) + 4\cos^2(n_1 k \Delta x)}, \qquad (8)$$

where $\xi = \cos\varphi \rho_{xx}^1 / n_1 \rho_0$, $n_1 \equiv \sqrt{\rho_{xx}^1 c_0^2 / B - (\rho_{xx}^1)^2 \sin^2\varphi / \rho_X \rho_Y}$ and $\Delta x = x_3 - x_0$. Here the reflectivity $R$ and transmissivity $T$ for $\alpha_i = -\alpha_{i'}$ can be shown to be identical to that of $\alpha_i = \alpha_{i'}$ (straight waveguides). Figure 1c gives the simulated transmissivity for $\alpha_1 = -\alpha_2 = \alpha_3 = 45°$ and $\alpha_1 = \alpha_2 = \alpha_3 = 45°$, which are in accordance with the theoretical analysis using Eq. 8. This suggests, once again, that the sharp corners of a specifically designed path bear virtually no influence on the sound reflectivity and transmissivity. Figure 1c also reveals that the EA-AMM can be excellently impedance-matched to the background medium for a wide range of angles of incidence (roughly -70° ≤ α ≤ 70°). This is in stark contrast with acoustic topological insulators and valley hall SCs, both of which require a specific angle of incidence in order to achieve momentum matching and efficient acoustic coupling.

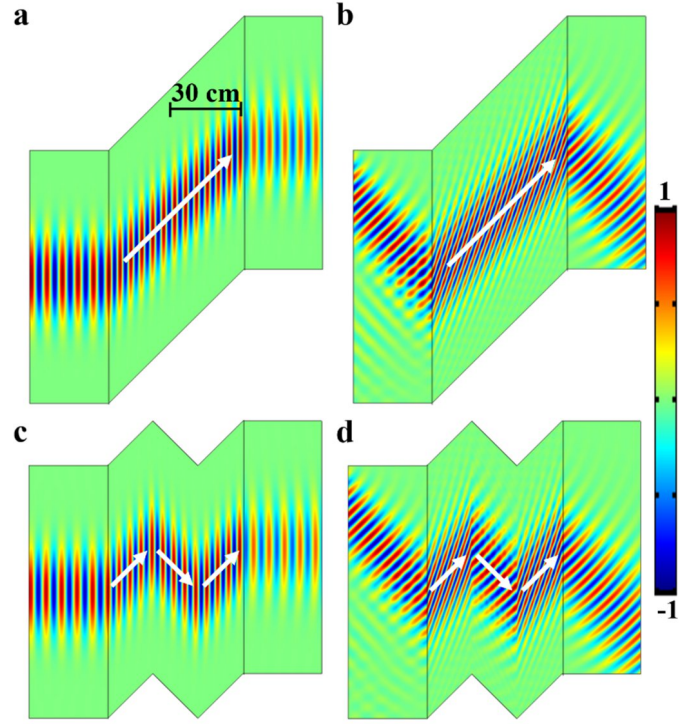

**Fig. 2 | Backscattering-immune propagation in zigzag paths and angle-dependent acoustic coupling. a-d** Simulated normalized pressure fields of EA-AMMs with the angles of rotation (**a**, **b**) $\alpha_1 = \alpha_2 = \alpha_3 = 45°$ and (**c**, **d**) $\alpha_1 = -\alpha_2 = \alpha_3 = 45°$ under (**a**, **c**) normal and (**b**, **d**) −45° incidence. The dimensions are $d$ = 4 mm, $a$ = 3.2 mm, $x_1$ = 200 mm, $x_2$ = 400 mm, and $x_3$ = 600 mm.

In addition, from the dispersion relation (Eq. S9, Supplementary Section 1), we can obtain the wavenumbers in the $x$ direction for the forward- and backward-going waves, i. e., $k_i^{\pm} = \pm n_i k - k \sin\varphi \, \rho_{xx}^i / \rho_{xy}^i$, which suggests that the phase velocities in regions $i$ and $i'$ are different with $\alpha_i = -\alpha_{i'}$ and $\varphi \neq 0$. However, when it comes to the normal incidence case ($\varphi = 0$), we have $k_i^{\pm} = \pm n_i k$, and therefore wave vectors (or phase velocities) with $\alpha_i = -\alpha_{i'}$ in regions $i$ and $i'$ are identical. Figure 2 shows the simulated acoustic fields for the EA-AMMs with $\alpha_1 = -\alpha_2 = \alpha_3 = 45°$ and $\alpha_1 = \alpha_2 = \alpha_3 = 45°$ at 5000 Hz, under normal and -45° incidence, respectively. The white arrows indicate the direction of energy flow, which is dictated by the angle of rotation only, as predicted by the theory. Diffraction is strongly suppressed in the metamaterial, as expected, due to the flat EFC derived from the strong density anisotropy[32]. Noteworthy is that the acoustic field shown in Fig. 2b mimics that of the negative refraction, which can be otherwise achieved by negative-index metamaterials, hyperbolic metamaterials, Parity-Time symmetric metasurfaces [48] or Weyl phononic crystals [49]. No appreciable backscattering is observed when sharp corners are present. It is important to point out that, the zigzag one-way propagation of sound in this EA-AMM is independent of the location where the wave impinges upon, since the EA-AMM is treated as a homogeneous medium. On the contrary, the edge state in acoustic topological insulators and valley hall SCs must be excited by a wave reaching the entrance of the waveguide created by the interface between two topologically distinct SCs. One implication of this is that a wide beam impinging upon a topologically protected waveguide will be largely reflected due to impedance mismatch. Finally, the phase velocities in neighboring regions are visually different for the oblique incidence case (Fig. 2b) and are identical for the normal incidence (Fig. 2a), confirming the earlier theoretical prediction.

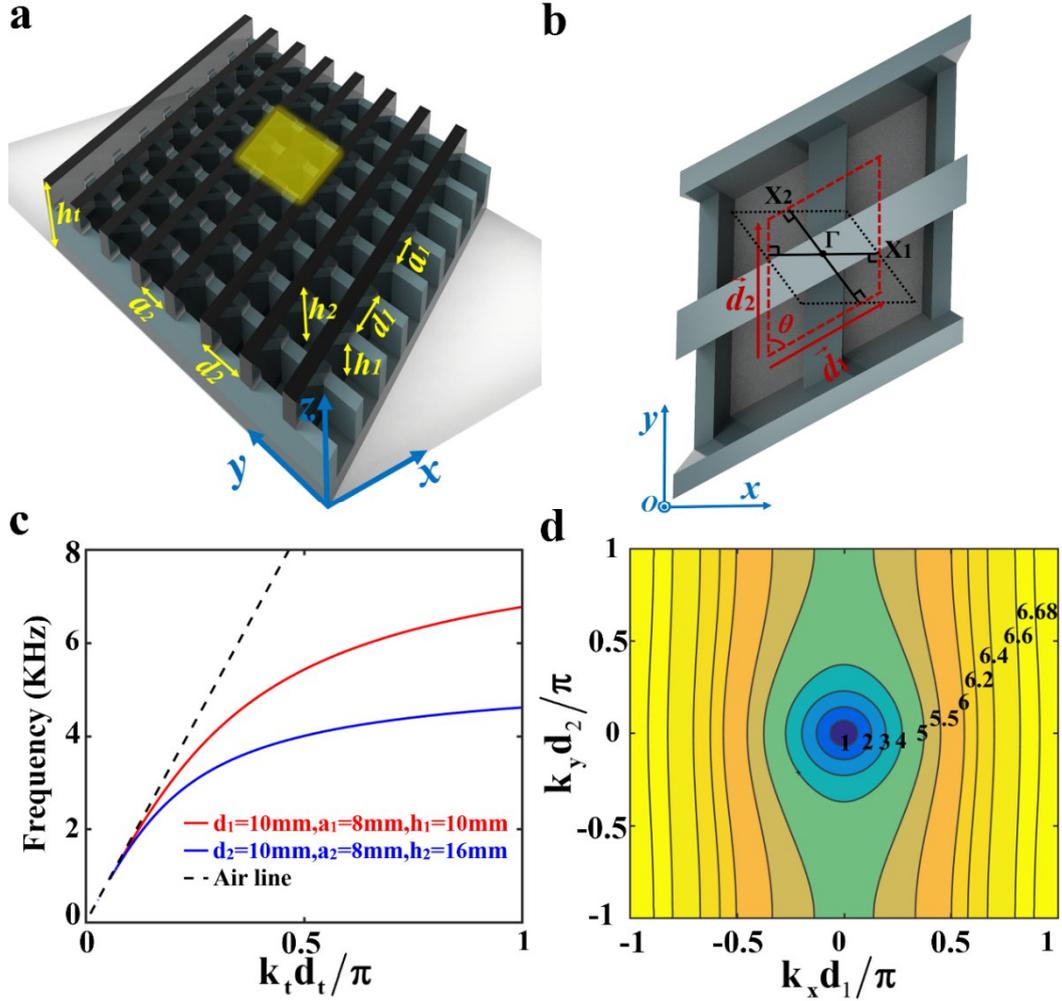

**Fig. 3 | Schematic illustration of the 2D corrugated PRB, dispersion relations of the 1D textured PRBs for manipulating SSAWs, and EFCs of the RS. a** The structure has heights $h_1$ and $h_2$ for the two arbitrary non-parallel directions and $h_t$ is the total height of the structure. The widths of the two grooves are $a_1 = 0.8d_1$ and $a_2 = 0.8d_2$, where $d_1$ and $d_2$ are the lattice constants of the unit cell. **b** An overhead view of part of the structure, which is marked yellow in (**a**). $\theta$ is the included angle of the two lattice vectors. The first Brillouin zone (FBZ) is depicted by the black dotted line. The unit cell is outlined by the red dashed line. **c** Dispersion relations of the two 1D textured PRB with different heights (bottom PRB in red and upper PRB in blue). **d** EFCs in FBZ of the RS. The frequency values are in kHz.

Though backscattering-immune sound propagation can be achieved by the foregoing 2D EA-AMM, challenges in full acoustic field mapping hinders further experimental exploration. Consequently, SSAWs are selected in lieu of bulk acoustic waves for validating this design concept. As shown in Fig. 3a, the proposed extremely anisotropic structure for manipulating SSAWs can be regarded as two 1D textured PRBs with different heights and orientations crossing from each other. As we will show later, the difference in the cutoff frequencies caused by the different heights for the two 1D textured PRBs leads to the existence of anisotropic forbidden band, which can efficiently suppress diffraction in order to produce collimation. The 1D textured PRB on the bottom (with the smaller height) plays an important role in confining the direction of SSAWs in the *xoy* plane. The other PRB on the top (with the larger height) acts as the 2D EA-AMM mentioned in the previous section, routing sound along predetermined paths. Figure 3b gives the overhead view of part of the structure, which is marked yellow in Fig. 3a. The unit cell is outlined by red dashed lines and the lattice constants are indicated in the figure. The lattice vector $\vec{d}_2$ of the unit cell is along the *y* direction, and the included angle between the two lattice vectors $\vec{d}_1$ and $\vec{d}_2$ is $\theta$. For the case $\theta = 0°$ or $180°$, the two periodic corrugated structures line up and the sample is reduced to a 1D textured PRB [35-39]. When $\theta = 90°$, the two lattice vectors are orthogonal and the overhead view of the sample manifests as a rectangle (or a square provided $d_1 = d_2$), thus it is denoted as a rectangular sample (RS). For the more general case of $\theta \neq 90°$, we have the parallelogram sample (PS). Here $\theta = 90° - \alpha$ ($\alpha$ is the angle of rotation) and we will adopt $\alpha$ hereinafter.

The dimensions of the structure in Fig. 3a are $d_1 = d_2 = 10$ mm, $a_1 = a_2 = 8$ mm, $h_1 = 10$ mm, and $h_2 = 16$ mm; $\alpha = 0°$ and $\alpha = \pm 45°$ are taken for further study. The cutoff frequencies of the two 1D textured PRBs are critical for analyzing the acoustical properties of this EA-AMM, which can be obtained by the dispersion relation of 1D SSAWs as follows [35]

$$\sqrt{k_{t,i}^2 - k_{o,i}^2}/k_{o,i} = \frac{a_i}{d_i}\tan(k_{o,i}h_i), \quad i = 1 \text{ or } 2, \tag{9}$$

where $k_{t,i}$ and $k_{o,i}$ are the propagating and operating wavenumbers, respectively. Figure 3c gives the dispersion curves. The corresponding cutoff frequencies can be readily calculated and are $f_1$= 6768 Hz and $f_2$= 4610 Hz, respectively.

EFCs can be further utilized to better understand the characteristics of the proposed SSAW-based EA-AMM. Here, RS is first taken for illustration. Theoretically, when the operating frequency $f_o$ satisfies $f_o < f_2 < f_1$, SSAWs propagate along both directions, and EFCs are elliptical due to the anisotropy of the material. When the operating frequency $f_o$ satisfies $f_2 < f_o < f_1$, only SSAWs propagating along the $x$ direction can be supported. To this end, the EFCs in FBZ of the RS are numerically calculated by COMSOL. As shown in Fig. 3d, the EFCs of the sample are elliptical at low frequencies, where the two 1D corrugated structures can both support propagating SSAWs, but have different group velocities due to different dispersion relations. As the frequency increases, the EFC transforms from being elliptical to approximately two flat lines, which is a direct manifestation of EA-AMMs. In this case, only the 1D structure along the $x$-direction can support the propagation of SSAWs and the direction of the energy flow is independent of the angle of incidence (Supplementary Section 4). Due to the flat EFCs, sound collimation is expected and is confirmed by simulations and measurements (Supplementary Section 5).

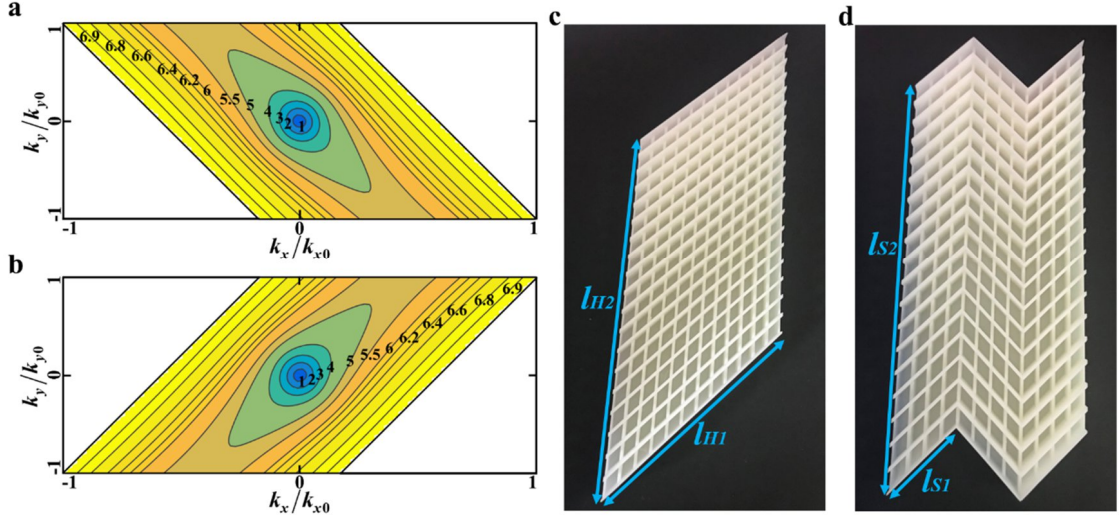

**Fig. 4 | EFCs and photos of the HPS and SPS. a-b** EFCs in FBZ of the PSs with (**a**) $\alpha = 45°$ and (**b**) $\alpha = -45°$, respectively. The frequencies are in kHz. **c-d** Photos of (**c**) the HPS and (**d**) SPS.

Sound collimation within a wide variety of paths can be created by PSs such as sloping and zigzag paths. Figure 4c shows a homogeneous PS (HPS) with a sloping path. We set $\alpha = 45°$, and in the two lattice vector directions, we have $l_{H1} = 12d_1$ and $l_{H2} = 19d_2$, respectively. A spliced PS (SPS) with a zigzag path is shown in Fig. 4d. It is constructed by combining three HPSs with successive angles of rotation $\alpha = 45°$, -45° and 45°. The three HPSs have the same length, i.e., $l_{S1} = l_{H1}/3$ and $l_{S2} = 19d_2$. These two samples are made of photopolymer plate with an array of periodic parallelogram holes, and the heights of the parallelogram hole in the two directions are different. The detailed experimental setup can be found in Supplementary Section 6.

For the PS, the FBZ can be obtained by the base vectors of reciprocal lattice in the *xoy* plane, i.e., $\vec{\beta}_1 = 2\pi/(d_1 \sin\theta)\hat{x}$ and $\vec{\beta}_2 = 2\pi/(d_2 \sin\theta)\hat{x}'$. $\hat{x}$ and $\hat{x}'$ are unit vectors in the $\Gamma X_1$ and $\Gamma X_2$ directions, which are perpendicular to $\vec{d}_2$ and $\vec{d}_1$, respectively (shown in Fig. 3b). Thus, the corresponding FBZ of the PSs with $\alpha = \pm 45°$ is the overlapped area of two different wave vectors with different ranges and directions, *i.e.*, $\beta_1 \in [-\sqrt{2}\pi/d_1, \sqrt{2}\pi/d_1]$ in the $\hat{x}$ direction and $\beta_2 \in [-\sqrt{2}\pi/d_2, \sqrt{2}\pi/d_2]$ in the $\hat{x}'$ direction, which is outlined by black dotted lines in Fig. 3b.

The EFCs in FBZ of the PSs for $\alpha = \pm 45°$ are given in Figs. 4a and b, respectively. Here $k_{x0} = \sqrt{2}\pi/d_1 + \pi/2d_2$ and $k_{y0} = \pi/2d_2$, which correspond to the maximum values of wavenumbers in the $x$ and $y$ directions. Similar to the RS, the flat EFCs at higher frequencies, which are parallel to $\Gamma X_2$, can be observed due to the extreme anisotropy of the AMM. Figures 5a and c give the simulated pressure fields of the HPS at two selected frequencies exhibiting flat EFCs: 6150 Hz and 6550 Hz. The results clearly demonstrate sound collimation. The energy is confined in a waveguide that is roughly 30 mm wide, which is only about half of the wavelength at 6150 Hz measured in free air, *i.e.*, 56 mm. This "subwavelength" feature of the waveguide can be attributed to the fact that the corrugated rigid surface significantly slows down the speed of sound[44]. Measured results, as shown in Figs. 5e and g, are in principle in good agreement. The slight difference is expected to be caused by sample defects and thermos-viscous losses, which are not considered in the simulation.

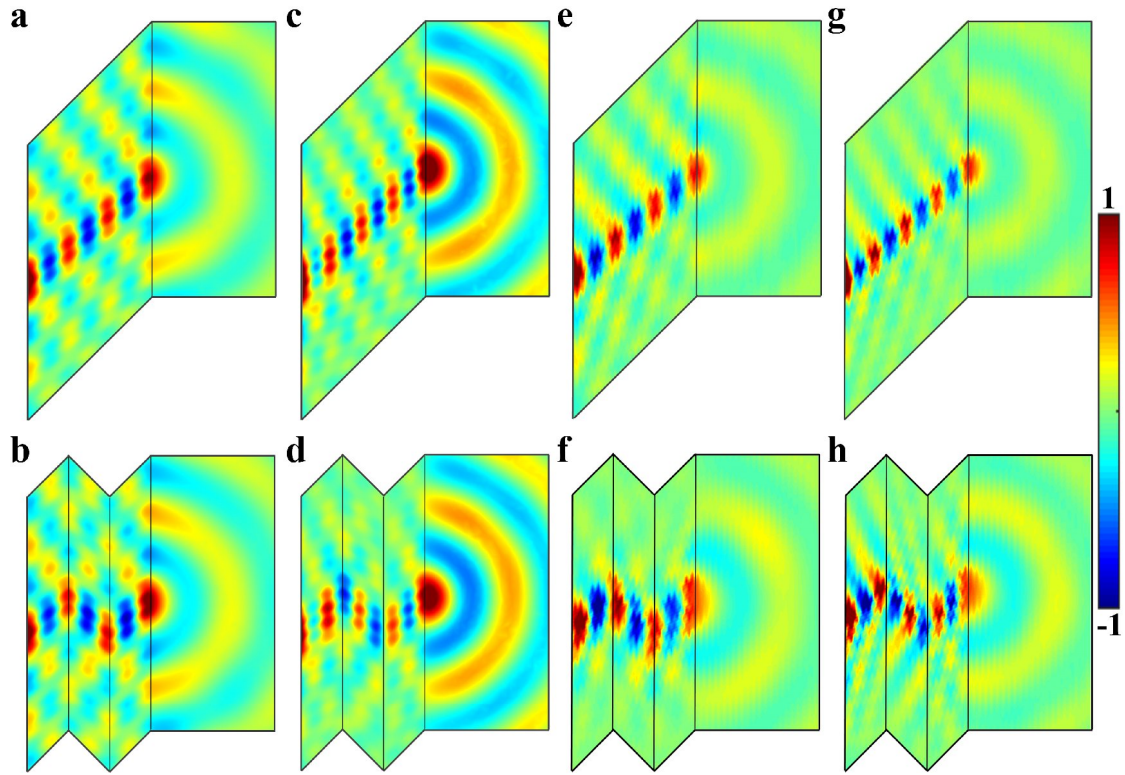

**Fig. 5 | Sound collimation with sloping and backscattering-immune zigzag paths.** (**a-d**) Simulated and (**e-h**) measured pressure field distribution for the domain 0.5 mm above the upper surface of the sample at different frequencies. **a-b**, **e-f** 6150 Hz; **c-d**, **g-h** 6550 Hz.

Remarkably, collimation in a backscattering-immune zigzag path can be observed with the SPS, which is illustrated in Figs. 5b, d, f and h, respectively, for simulations and measurements. Such a phenomenon can be explained as follows: first, the extreme anisotropy of the textured surface engenders collimation of SSAWs so that the energy is strongly confined and travels along the predetermined path. Second, the judiciously designed angles of rotation ensure that the backscattering at the corners can be perfectly suppressed, as predicted by the theory established by this work. To further prove that there is no backscattering at the sharp corners, Figure 6 compares the transmitted sound energy between the HPS (no bending) and SPS (with bending). Notice that the two waveguides have the identical length and virtually the same sound transmission can be seen in Fig.6 even though the SPS contains sharp corners. One important observation is that the size of the wave-guiding EA-AMM is very small relative to the wavelength in free air; for example, the width of the EA-AMM $l_{H2}/l_{S2}$ is merely 3.4 wavelengths at 6150 Hz. This again can be compared with acoustic topological insulators that typically have a width on the order of 10 wavelengths [14,17]. Additional simulations show that $l_{S2}$ can be reduced by more than a half ($l_{S2}$=1.6 wavelength) and the resulting EA-AMM would still yield excellent performance (Supplementary Section 7). Therefore, the proposed metamaterial would have an inherent advantage over SCs possessing topological edge states for building portable/miniature devices. Finally, we point out two important facts: (a) sloping and zigzag paths with arbitrary angles are feasible (Supplementary Section 7) using the current design scheme; (2) as predicted by the theory, this EA-AMM can be efficiently coupled with acoustic sources almost arbitrarily located along the vertical direction, offering the possibility that the sound propagation path can be tuned flexibly by adjusting the source-position. This is generally not possible with acoustic topological insulators or valley hall SCs (Supplementary Section 7).

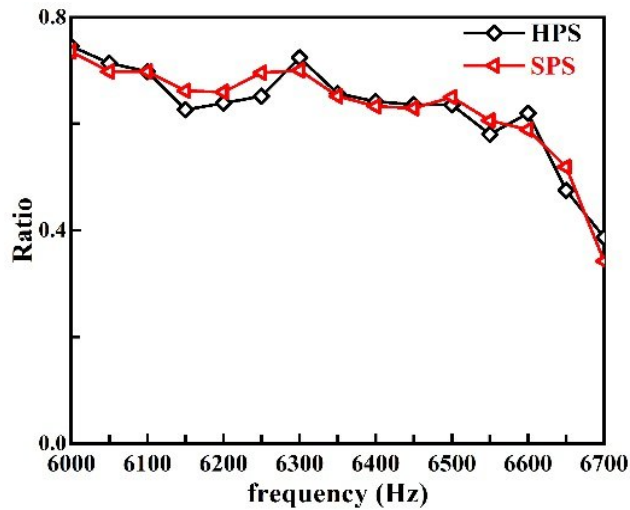

**Fig. 6 | Ratio between the measured energy at the exit of the waveguide and the measured energy at the entrance of the waveguide for both HPS and SPS.**

In summary, we have devised and experimentally demonstrated a new class of EA-AMM that support backscattering-immune wave propagation around sharp corners. Based on the wave/Helmholtz equation and homogenization theory, we theoretically derived the exact conditions for perfect suppression of backscattering in zigzag paths with sharp corners. Our proposed wave guiding strategy hinges on the usage of metamaterials exhibiting strong anisotropy, and is numerically and experimentally shown to support robust one-way propagation in zigzag paths without relying on topological edge states or defect modes. While this study focuses on sawtooth-like zigzag waveguides, other types of waveguides with sharp corners can also be designed (Supplementary Section 8). When compared to acoustic topological insulators and valley hall SCs, the proposed EA-AMM is superior in several aspects such as the smaller size and more efficient acoustic coupling. One limitation of our design, though, is that the bending angles at the corners have to be identical. This is, however, only a limitation for the SSAW AMM design, and does not apply to the effective medium and bulk acoustic wave AMMs (Supplementary Section 8). Finally, the theory established in this work is generic and can be extended to electromagnetic waves (Supplementary Section 9). Our proposed method could be useful for many applications such as delay line design, noise control, acoustic cloaking, and on-chip wave manipulation.

## Acknowledgements

This work was supported by the National Key Research and Development Program of

China (2017YFA0700201, 2017YFA0700202, 2017YFA0700203), the National Natural Science Foundation of China (61722106 and 61731010), the 111 Project (Grant No. 111-2-05), and the Postgraduate Research & Practice Innovation Program of Jiangsu Province (KYCX18_0099).

**Author contributions**

Q. C. conceived the idea for this work. Q. C., Y. J. and T. J. C. suggested the designs, planned and supervised the work. L. W. carried out the analytical modeling and designed the samples. L. W., M. O., W. C. and H. J. conducted the theoretical calculation. L. W., C. Z., J. K., J. Y. and Y. D. performed the experiments and data analysis. All authors participated in drafting the manuscript, discussion and interpretation of the data.

**Competing financial interests**

The authors declare no competing financial interests.